\documentstyle[11pt,paspconf]{article}

\begin{document}

\title{Heavy Bars and Light Spirals --
Taking Advantage of Asymmetries in Galaxies}
\author{A. C. Quillen}
\affil{Steward Observatory, University of Arizona}

\begin{abstract}
I summarize a series of studies where the underlying
theme is to exploit non-axisymmetry disk structures 
such as bars and spiral arms.  By measuring
the strength of gravitational forces from these
structures we can measure the mass in them.
The advantage of this approach is that mass to light
ratios can be constrained in a way independent
of the dark matter profile.  By examining a variety
of types of galaxies it appears that a correlation between
morphology and dark matter fraction emerges.

\end{abstract}
\section{Examples of Using Assymmetries in Galaxies to Constrain
the Disk Mass}

Axisymmetric fits to rotation curves in disk galaxies and velocity
profiles in elliptical galaxies allow a large
range of freedom in choosing the mass to light (M/L) ratio of 
the stellar population.  This is because there is little
observational constraint on the actual form of the dark matter profile.
Here we introduce a different approach.  By measuring the strength
of non-axisymmetric structures we can constrain M/L 
in disk galaxies in a way that is independent of any
assumptions about the dark matter profile.

\subsection{Resonant Orbits Outside a Bar} 

In Quillen \& Frogel 1997 we studied the affect of the mass 
of the bar on the shape of the resonance R1 orbits at the Outer
Lindblad resonance.  By modeling the shape
and velocity field of the outer ring in NGC 6782 we found that 
the mass of the bar must be quite massive, 
nearly that of a maximal disk (implying little dark matter within the bar).

\subsection{Gas response to spiral structure} 
A strong spiral gas response results from a gravitational perturbation
strong enough to produce shocks.  When strong spiral
structure is observed in HI we can then place a lower limit on
M/L.  If velocity perturbations are small
then there the mass in spiral structure cannot be large.
This lets us place an upper limit on M/L.
In Quillen \& Pickering 1997 
we place upper and lower limits on M/L of the spiral arms
in two low surface brightness galaxies.
We find that they have normal stellar populations.
Since the current star formation rates of these galaxies
is very low they must have had much higher star formation
rates previously.  These galaxies are therefore good examples
of old but {\bf faded} galaxies.
This technique can also be used to infer the presence
of a dim or quiescent stellar disk from the HI morphology 
(e.g. Quillen 1998).

\subsection{Moderate Redshift Galaxies} 

In Quillen \& Sarajedini 1998, 
we consider distant galaxies observed by HST which have
clearly evident spiral structure.
Since they display spiral structure they must have disks
that are cold and massive enough to support and respond
to spiral density waves and yet not fragment.
We use the Toomre Q parameter to place both lower and upper
limits on the ratio of the mass to light ratio to 
the stellar velocity dispersion, our two unknowns.

The maximal disk mass to light ratio places an upper limit
on M/L of these galaxies.  
We find that they have low maximal disk
mass to light ratios and so are probably young with elevated
star formation rates compared to local galaxies.   
The most likely values for the mass
to light ratios and velocity dispersions suggest that they
are dark matter dominated.

\subsection{When Isochromes Differ from Isophotes} 

In Quillen et al. 1996 we showed that 
in a disk galaxy with both a radial color gradient and non-circular
motion, isochromes or iso-color contours
should follow the shape of closed stellar orbits,
and the ellipticity of the isophotes should vary as a function
of wavelength.
While we initially proposed this model to search 
for triaxial galaxy halos, this phenomenon is observed
in barred galaxies and could be used to place constraints on the
orbit distribution in them.

\section{A Pattern?}

Our ring galaxy work supports
studies (either of gas dynamics or of dynamical friction) 
by Sellwood and collaborators that also suggest that early type
barred galaxies are likely to be nearly maximal disk.
This would suggest that high surface brightness systems with
high stellar densities have low dark matter fractions.
On the other hand our constraints on M/L 
for the distant spiral galaxies suggest that
these galaxies have somewhat submaximal disks.
The low surface brightness, though they have normal
stellar populations are highly dark matter dominated.
This suggests that a `dark matter' conspiracy, 
(where the dark matter profile depends on the history
of merging and star-formation in the galaxy) rather
than dark matter domination at all radii, is a more likely
explanation for the Tully-Fisher relation.

\end{document}